\documentclass{elsart}

\usepackage{epsfig}
\usepackage{amssymb}

\newcommand{\proto}{0.1~km$^2$}
\newcommand{\med}{Mediterranean Sea}
\begin{document}

\begin{frontmatter}


%
\title{The ANTARES Optical Module}
{\bf The ANTARES Collaboration}\\
%

\author[lam]{P.~Amram},
\author[dipartimento-genova]{M.~Anghinolfi},
\author[dapnia]{S.~Anvar},
\author[dapnia]{F.E.~Ardellier-Desages},
\author[cppm]{E.~Aslanides},
\author[cppm]{J-J.~Aubert},
\author[dapnia]{R.~Azoulay},
\author[oxford]{D.~Bailey},
\author[cppm]{S.~Basa},
\author[dipartimento-genova]{M.~Battaglieri},
% ...
\address[cppm]{CPPM -- Centre de Physique des Particules de Marseille, CNRS/IN2P3
  Universit\'e de la M\'editerran\'ee Aix-Marseille II, Marseille, France}
% ...
\address[dapnia]{DAPNIA - CEA/DSM, Saclay, France}
\address[dipartimento-bari]{Dipartimento Interateneo di Fisica e Sezione INFN, Bari, Italy}
\address[dipartimento-bologna]{Dipartimento di Fisica dell'Universit\`a e Sezione INFN, Bologna, Italy}
\address[dipartimento-catania]{Dipartimento di Fisica ed Astronomia dell'Universit\`a e Sezione INFN, Catania, Italy}
\address[dipartimento-genova]{Dipartimento di Fisica dell'Universit\`a e Sezione INFN, Genova, Italy}
\address[dipartimento-roma]{Dipartimento di Fisica dell'Universit\`a "La Sapienza" e Sezione INFN, Roma, Italy}
% ...
\address[lam]{LAM -- Laboratoire d'Astronomie Marseille, CNRS/INSU Universit\'e de Provence Aix-Marseille I, France}
% ...
\address[oxford]{University of Oxford, Department of Physics, Oxford, 
  United Kingdom}
% ...
%

%
%

\begin{abstract}
The ANTARES collaboration is building a deep sea neutrino telescope in the Mediterranean Sea. This detector will cover a sensitive area of typically 
0.1~km$^2$ and will be equipped with about 1000 optical modules. Each of these optical modules consists of a large area photomultiplier and its associated electronics housed in a pressure resistant glass sphere. The design of the ANTARES optical module, which is a key element of the detector,
has been finalized following extensive R\&D studies and is reviewed here in detail.
\end{abstract}

\end{frontmatter}

\section{Introduction}
\label{intro}

Neutrinos offer a unique opportunity to explore the Universe in depth over a wide energy range~\cite{bib:general}. 
However, since neutrino fluxes at high energy (E$_\nu>$~TeV) are expected to be very low~\cite{bib:theoflux}, a very large detector volume is required. A detector immersed in the sea provides a cheap and
efficient method of observing high energy muon neutrinos by detecting the muon produced from a charged current interaction in the matter surrounding the detector. The muon emits \v{C}erenkov light as it passes through the water, and this  light can be detected by a three-dimensional array of optical sensors called Optical Modules (OM). 
The measurement of the arrival time of the \v{C}erenkov light at each OM allows the reconstruction of the muon direction, and the amount of light collected can be used to estimate the muon energy.  

The ANTARES collaboration has started the construction of a \proto\ detector to be immersed at 2400~m depth
in the \med\ 40~km off the French coast (42$^0$ 50' N, 6$^0$ 10' E). 
This detector will be equipped with about 1000 OMs and has been 
designed to be competitive with other such detectors 
(DUMAND~\cite{bib:dumand}, Baikal~\cite{bib:baikal}, AMANDA~\cite{bib:amanda}
and NESTOR~\cite{bib:nestor}),
notably in terms of angular
resolution. To reach this goal, extensive R\&D studies have been carried out during the past years
in order to optimize the design of each element of the telescope, particularly its most fundamental component, the optical module. This optimization work
has involved  
a variety of measurements, both in the laboratory and {\it in situ}, 
as well as Monte Carlo simulations. 
This paper gives the results of these studies and describes in detail the final design of the ANTARES optical module.

In section~\ref{sec:omspecif}, the constraints inherent to the operation of a 
deep sea neutrino detector are reviewed and the specifications for the optical modules are listed. 
Section~\ref{sec:omassembly} describes the main features of the different components of the OM as well as the assembly procedure.
The test set-ups and the main results of the tests 
performed by the collaboration during the past years are presented  
in section~\ref{sec:omtest}. Finally, section~\ref{sec:massprod}
deals with mass production and quality control aspects.

\section{Constraints  for a deep-sea neutrino telescope} 
\label{sec:omspecif}


The detection of high energy cosmic neutrinos together
with the expected background sources impose particular constraints on
the design of a deep undersea detector and on its basic element, the
optical module.

\subsection{Background sources}


Atmospheric muons, originating from mesons
which are copiously produced by the interaction of
cosmic rays with the Earth's atmosphere, are the dominant source of background
for a neutrino telescope based on the detection of the muons produced in  neutrino interactions. Atmospheric muons are about 10 orders of magnitude  
more numerous at sea level than muons produced by neutrinos. Shielding the telescope against atmospheric muons is thus one of the most important 
considerations of the telescope design.

A straightforward way of reducing the down-going atmospheric muon flux 
is to immerse the 
telescope as deep as possible in the sea. At 2400~m, the depth chosen for the ANTARES 
detector, the atmospheric muon flux is already reduced by more than 4 orders of magnitude. 
Though significant, this is still insufficient for the 
measurement of down-going neutrino  interactions and a more radical method is to observe 
only up-going muons, using the Earth itself as a natural shield.
Since the material of the Earth stops all up-going atmospheric muons, 
the telescope is only sensitive to muons produced in neutrino interactions.

Important sources of background light come from natural radioactivity of sea
water ($^{40}$K decays) and from the presence of bioluminescent organisms.
This optical background has both continuous and burst components which have
been studied in detail at the ANTARES site~\cite{bib:nathalie}.
Though it has no significant impact on the OM design itself, 
it does constrain the trigger scheme as well as the  data acquisition system.



 The expected signal fluxes are very low, so the surface area of the
detector must be as large as possible, preferably \hbox{1 km$^2$} or more. 
The 0.1~km$^2$ telescope foreseen as the first phase of the ANTARES
project, despite its limited size,  should provide some initial
results on high energy neutrino sources.

\subsection{The ANTARES project}

\begin{figure}
\vspace{-2cm}
 \begin{center}
 \mbox{\epsfig{file=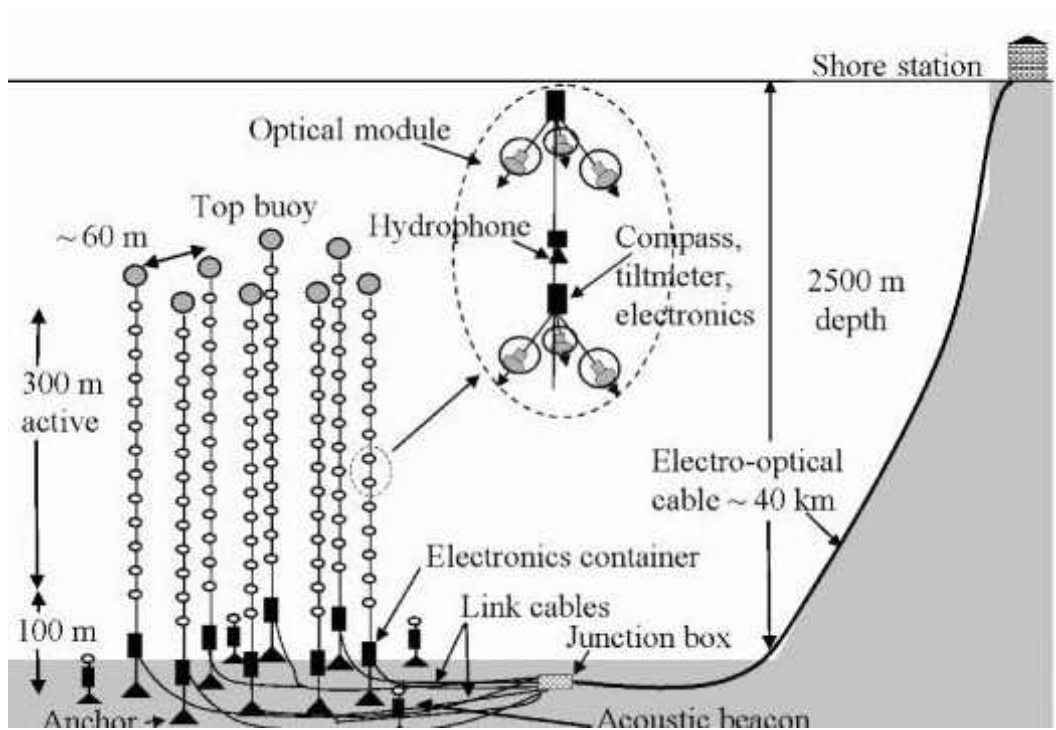,width=1\linewidth}}
 \end{center}
 \vspace{1.5cm}
 \caption{Sketch of the ANTARES detector. Each string is about 400 metres long and is equipped with 90 optical modules 
grouped together in triplets at 30 storeys.}
 \label{fig:detecteur}
\end{figure}

\begin{figure}
\vspace{-0cm}
 \begin{center}
 \mbox{\epsfig{file=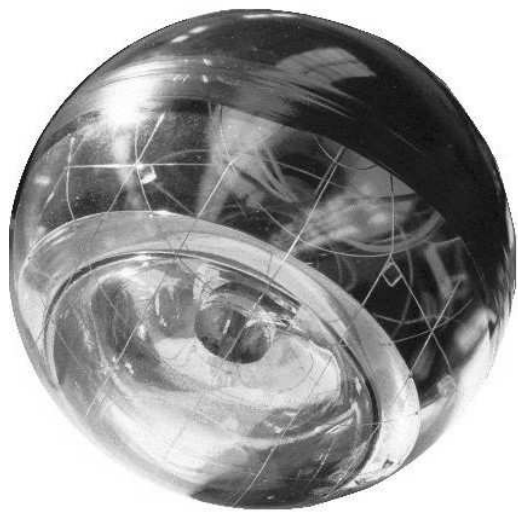,width=0.5\linewidth}}
 \end{center}
 \vspace{0cm}
 \caption{Photograph of an optical module.}
 \label{fig:photoOM}
\end{figure}

The ANTARES detector, shown in figure \ref{fig:detecteur}, 
consists  of a three-dimensional array of optical modules supported by  flexible vertical strings anchored to the sea bottom~\cite{bib:refAntares}.
 Each OM (figure~\ref{fig:photoOM}) contains a large area photosensor (photomultiplier) protected from the outside pressure. 
The detector is connected to the shore by an electro-optical cable for data transmission and power supply. 

\begin{figure}
\vspace{-0cm}
 \begin{center}
 \mbox{\epsfig{file=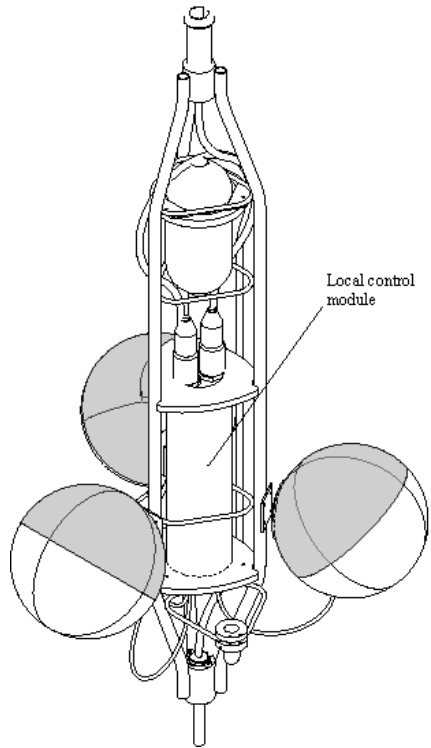,width=0.4\linewidth}}
 \end{center}
 \vspace{0cm}
 \caption{Drawing of a storey equipped with 3 OMs forming a triplet.
 OMs are positioned with the axis of the photomultiplier tubes pointing
 at 45$^\circ$ below the horizontal plane. They are 
 connected to a local control module located at the centre of the storey.}
 \label{fig:omf}
\end{figure}

These strings are about 400~m long and separated from each other by 
at least 60~m. 
Each string is equipped with 90 OMs,
grouped together in triplets at 30 storeys (figure \ref{fig:omf}) and connected to a local control
module which holds the electronics for the trigger, signal digitization, 
slow-control
and data transmission. The storeys on a given string are spaced 12~m vertically
along an interconnecting elctro-opto-mechanical cable.
OMs at each storey are positioned with the axis of the photomultiplier tubes (PMTs) pointing outwards and inclined at 45$^\circ$ below the horizontal. 

The strings must be flexible for deployment. Deep sea currents may modify the shape of the strings, so an accurate positioning system 
is required to monitor their deformation during and after deployment and to measure the position of each element with a precision better than 20~cm. Several additional instruments are needed to monitor the environmental conditions
and to calibrate the detector in time and in energy. 

\subsection{Optical module requirements}

The scientific goals and the environmental constraints on the ANTARES detector described 
in the previous sections lead to the following global requirements for the OM:
\begin{itemize}  
\item 
Light detection must be optimized. A hemispherical photomultiplier with a large photocathode surface must be used. For maximum detection efficiency, 
special care must be taken to ensure the best possible optical coupling between the water and the photocathode, and the influence of the Earth's magnetic field must be minimized.
\item 
 Due to the pressure at a depth of 2400~m, the photomultiplier and its associated electronics must be housed in a pressure-resistant glass sphere. The OM must withstand conditions likely to occur during  sea operations 
(shocks, corrosion, vibrations, exposure to sunlight, etc.). 
\item 
The OM life-time should be greater than 10 years. 
Due to the deep-sea environment, it will be difficult and expensive to repair the detector. Therefore, the reliability of all components should be high.
\end{itemize}


\section{The optical module}
\label{sec:omassembly}

\begin{figure}
\vspace{-0cm}
 \begin{center}
 \mbox{\epsfig{file=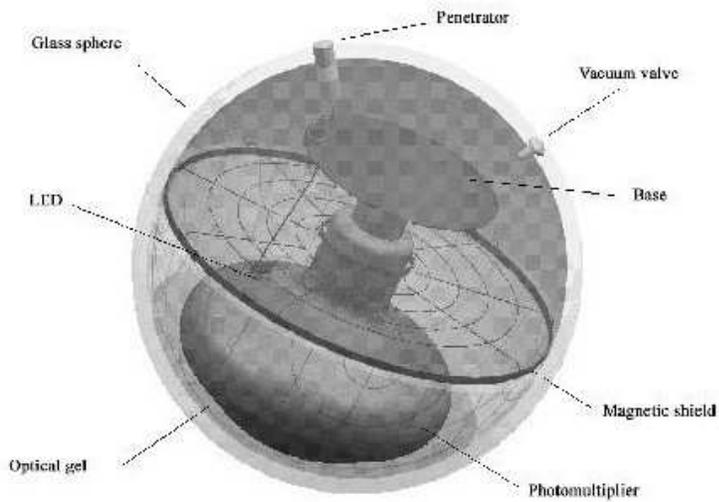,width=0.75\linewidth}}
 \end{center}
 \vspace{0cm}
 \caption{Schematic 3D view of the ANTARES optical module 
and its components.}
 \label{fig:OM_layout}
\end{figure}

The layout of the  OM is shown in figure~\ref{fig:OM_layout}. 
Its main component is a large area 
hemispherical photomultiplier (PMT) glued in a pressure resistant 
glass sphere with optical gel. 
A $\mu$-metal cage is used to shield the PMT against the Earth's magnetic field.
Electronics inside the OM are reduced to a minimum, namely: the PMT high voltage power supply and a LED system used for internal calibration. The specifications for all the components 
are described in this section as well as the main technical solutions adopted.

\subsection{Component description}

\subsubsection{Photomultiplier}         
 The reconstruction of the neutrino direction is based on 
the measurement of the arrival time of 
the \v{C}erenkov light at each OM. Optimization studies based on 
computer simulations have shown that,  
to achieve the desired detector performance, 
the photomultiplier must have a large photocathode surface ($\ge$~500~cm$^2$), 
timing resolution better than 3.0~ns (FWHM), 
gain $G$ greater than 5$\cdot$10$^7$ and good energy response:  
peak-to-valley ratio P/V $\ge$2 and single photo-electron resolution
$\sigma_E/E \le 50$\%. 

\begin{figure}
\vspace{-0cm}
 \begin{center}
 \mbox{\epsfig{file=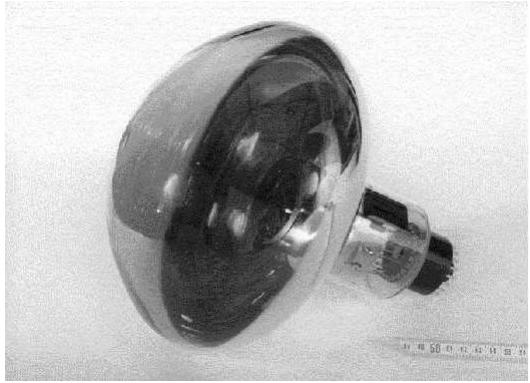,width=0.5\linewidth}}
 \end{center}
 \vspace{0cm}
 \caption{A 10$"$ photomultiplier model R7081-20 from Hamamatsu.}
 \label{fig:pmt}
\end{figure}

Seven types of photomultipliers with various photocathode diameters 
from three different manufacturers have
been  evaluated: 8$"$ (R5912-02), 10$"$ (R7081-20) and 13$"$ (R8055) from Hamamatsu\footnote{Hamamatsu Photonics, 812 joko-cho, Hamamatsu city, 431-31 Japan.}, 9$"$ (XP1802) and 10.6$"$ (XP1804/D2) from Photonis\footnote{
Photonis, Avenue Roger Roncier, Z.I. Beauregard
B.P. 520, 19106 Brive, France.},
8$"$ (ETL9353) and 11$"$ (D694) from Electron Tubes Ltd\footnote{
Electron Tubes Limited, Bury Street, Ruislip, Middlesex, HA4 7TA, England.}. 
A figure of merit was evaluated for each of these products. 
The photomultiplier selected is the 14-stage 10$"$ Hamamatsu 
photomultiplier R7081-20
(figure~\ref{fig:pmt}). Its main characteristics are summarized in
table~\ref{tab:pmt}. Details of the PMT 
measurements will be reported 
separately~\cite{bib:pm_ANTARES}.

\begin{table}
\begin{center}
\caption{Average properties of the 10$"$ Hamamatsu 
photomultiplier R7081-20. 
The transit time spread (TTS), the peak-to-valley ratio (P/V), and 
the resolution  $\sigma_E/E$ are measured from 
the single photo-electron spectrum. 
The dark count (DC) rate is measured at room temperature and at 
0.25 photo-electron level. 
Quoted values are for a nominal gain of 5$\cdot10^7$.}
\begin{tabular}{lc}
  \hline
  Photocathode area     &  500 cm$^2$ \\
  High Voltage          & 1760 V \\ 
  TTS (FWHM)            & 2.6 ns \\
  P/V                   &  2.7  \\  
 $\sigma_E/E$           &   40 \% \\ 
DC rate                 & $\sim 1900$   Hz      \\
  \hline
\end{tabular}
\label{tab:pmt}
\end{center}
\end{table}

\subsubsection{High-pressure container}

The photomultiplier is enclosed in a glass sphere to protect it from the
pressure of the surrounding water while ensuring good light transmission. 
The main requirements for the high pressure glass sphere to be used 
for the OM have been identified as:
\begin{itemize}
\item the ability to resist high hydrostatic pressures: about 260~bars during normal 
operation 
and up to 700~bars  in qualification tests; 
\item  transparency to photons in the wavelength range 400-500~nm;
\item  watertightness, at both low and high pressure; 
\item electrical communication with the outside via a suitable
               watertight and pressure-resistant connection
(signal output and  control of the photomultiplier high voltage and LED system);  
\item optical matching, implying a refractive index close to that of sea water and of the PMT glass window.
\end{itemize}

\begin{figure}
\vspace{-0cm}
 \begin{center}
 \mbox{\epsfig{file=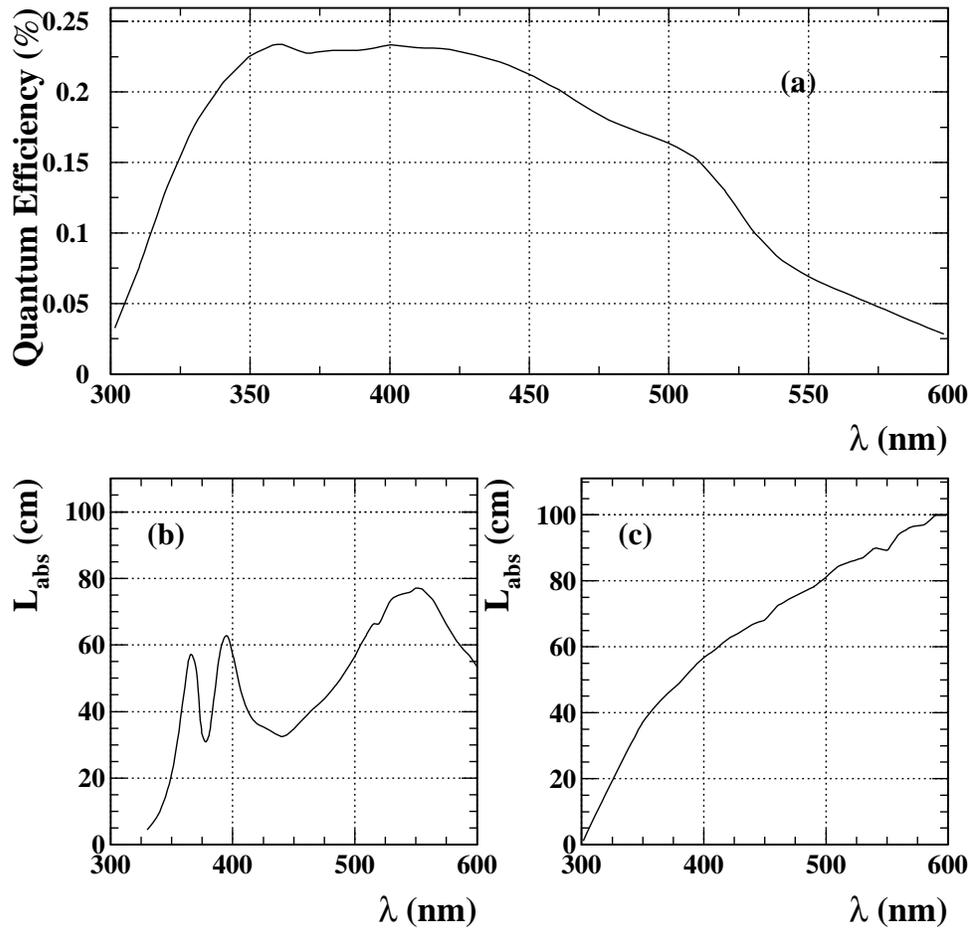,width=1\linewidth}}
 \end{center}
 \vspace{0cm}
 \caption{(a) PMT quantum efficiency (from Hamamatsu), (b) measured absorption length of the glass sphere  and (c) of the silicone gel  as a function of the incident light wavelength.}
 \label{fig:labs}
\end{figure}

The glass sphere chosen is a commercial product from  
Nautilus\footnote{Nautilus Marine Service GmbH, 
Heferwende 3, D-28357 Bremen, Germany.} with proven performance for deep ocean instrument housings 
(table~\ref{tab:nautilus}).  
As shown in figure~\ref{fig:labs}, the attenuation length is greater than 
30~cm above 350~nm, which corresponds to a transmission of at least 95\%. 
The sphere is fully resistant to corrosion; 
it is chemically, electrically and magnetically inert. 
Each sphere is made of two hemispherical parts, and
watertightness is ensured by precise grinding of the interface plane. 
The lower hemisphere (facing the PMT photocathode) supports the 
photomultiplier and the magnetic shielding, glued in place with silicone gel.
The upper part is 
painted black  and houses a penetrator, which provides the electrical connection between the inside and the outside, 
and a vacuum valve, which is used to set and monitor the OM internal pressure. 
The vacuum valve is the only metallic part of the glass sphere: it is made of titanium to prevent any risk of corrosion.

\begin{table}\begin{center}
\caption{Main properties of the high-pressure glass sphere.}
~\\
\begin{tabular}{ll} \hline
 Material & Vitrovex  8330 \\
 & (low-activity borosilicate glass) \\ 
 Outer diameter &  432 mm \\  
Thickness &  15 mm minimum \\  
Refractive index &  1.47 ($300<\lambda<600$~nm) \\ 
 Transmission & $>$95\% above 350~nm \\  
Depth rating   & 6700~m \\
\hline
\end{tabular}
\label{tab:nautilus}
\end{center}
\end{table}
 

\subsubsection{Magnetic shielding}

The Earth's 
magnetic field modifies the electron trajectory in the PMT, 
especially 
between the photocathode and the first  dynode, and degrades the uniformity of 
the response. At the ANTARES site, the ambient magnetic field is expected to be uniform,
pointing downward at 23$^\circ$ from the vertical with an amplitude 
of about 44~$\mu$T.
Laboratory measurements have shown that such a magnetic field degrades
the transit time spread and peak-to-valley ratio
of the 10$"$ Hamamatsu PMT operating in nominal conditions by up to 30\% 
(depending on the orientation).

A magnetic shield is thus used to make the PMT response sufficiently
homogeneous over the photocathode surface, 
while minimizing shadowing effects.
For this purpose, 
a wire cage made of $\mu$-metal, a nickel-iron
alloy 
with very high magnetic permeability (between 50000 and 150000) 
at low field strengths from  Sprint Metal\footnote{Sprint Metal, Groupe Usinor, 58160 Imphy, France.}, has been developed and tested.

\begin{figure}
\vspace{-0cm}
 \begin{center}
 \mbox{\epsfig{file=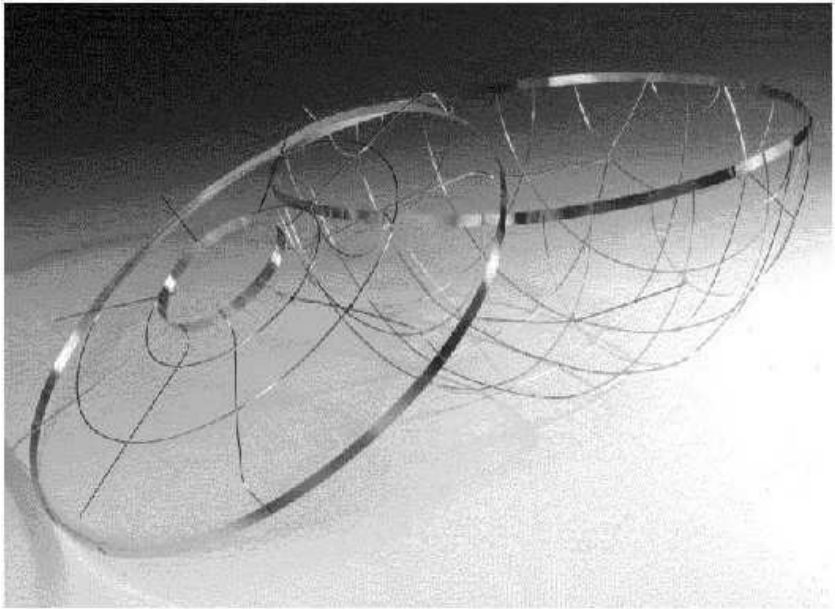,width=0.8\linewidth}}
 \end{center}
 \vspace{0cm}
 \caption{$\mu$-metal cage for magnetic shielding.
 The hemispherical part has    
outer diameter $\Phi$=395~mm and height H=199~mm. The flat part has  $\Phi=$~399 mm, H=30~mm. The two parts are linked together with four clamps of $\mu$-metal wire.}
 \label{fig:mucage}
\end{figure}

As shown in figure~\ref{fig:mucage}, the cage is composed of 2 parts:
a hemispherical part which covers the entire photocathode of 
the PMT, and a flat part 
consisting of a disk with a hole in its centre to allow the neck of the PMT to fit through.
It is made out of $\mu$-metal wire (1.1~mm diameter) clamped together at 
the interface of the two parts. 
Since $\mu$-metal loses its properties when mechanically stressed, 
the cages are
constructed in two steps: first the $\mu$-metal  wires are 
point welded 
to obtain the required configuration, then the cages are cleaned and baked at 1070$^\circ$C  
for 3 hours to restore the initial properties by annealing.

The pitch of the grid (68~mm $\times$ 68~mm) was chosen to minimize the loss 
due to its own shadow on the photocathode (less than 4\%) while reducing the ambient magnetic
field by more than a factor of 2.5 everywhere inside the volume of the cage.
In addition, the PMT in the detector will be fixed
such  that the dynodes are horizontal. Assuming a vertical
direction for
the residual ambiant field, this orientation minimizes the gain
change
between several orientations of the storey around the vertical.

\subsubsection{Optical glue}

The  optical glue serves two purposes. It ensures an optical 
link between the glass sphere and the PMT photocathode, and it fixes the mechanical position of the different elements (PMT and $\mu$-metal cage)
inside the OM.

The main requirements for the selection of the optical glue were 
the following:
\begin{itemize}
\item it should be highly transparent and have a refractive index as 
close as possible to that of the glass envelope and the PMT window 
(to reduce Fresnel reflection);
\item it should be firm enough to hold the different components together and,
 at the same time, sufficiently
 elastic to absorb shocks and vibrations during
transportation and deployment, as well as to absorb 
the deformation of the glass sphere 
under pressure (a reduction of 1.2~mm in diameter at a depth of 2400~m);
\item its optical and mechanical properties should be sufficiently
stable over a  10-year period.
\end{itemize}

The material chosen is a silicone rubber  
gel (reference SilGel 612 A/B) from 
Wacker\footnote{Wacker-Chemie GmbH, Hans-Seidel-Platz 4, 81737 Munich, Germany.}.
It is a highly transparent 
gel which needs 4 hours to polymerize at 23$^\circ$C. It offers pronounced tackiness and very good mechanical damping properties.
Details of the gluing procedure, important for optimal use of this gel,  
are given in subsection~\ref{ref:assembly}. After polymerization, 
the gel has a 
refractive index of 1.404, and an attenuation length  
of about 60~cm above 400~nm, increasing with wavelength 
(figure~\ref{fig:labs}).

\subsubsection{High voltage system}

The role of the high voltage system is to provide the PMT with 
all intermediate 
voltages necessary for correct operation. This consists of a constant 
focussing voltage 
between the photocathode and the first dynode and a variable amplification 
voltage to be applied
between the first dynode, the intermediate dynodes and the anode. 
Our tests have shown that the optimal  PMT response
was obtained with a focusing voltage of 800~V, while the amplification voltage
should vary between 500~V and 1500~V to allow for gain adjustment over  a reasonable range. 
The main requirements of the HV system are:
\begin{itemize}
\item low power consumption
\item high stability at constant average light flash rate
\item acceptable stability and short recovery time for strong light pulses
and  periods of high rate
\item long term reliability
\item compactness
\end{itemize}

\begin{figure}
\vspace{-0cm}
 \begin{center}
 \mbox{\epsfig{file=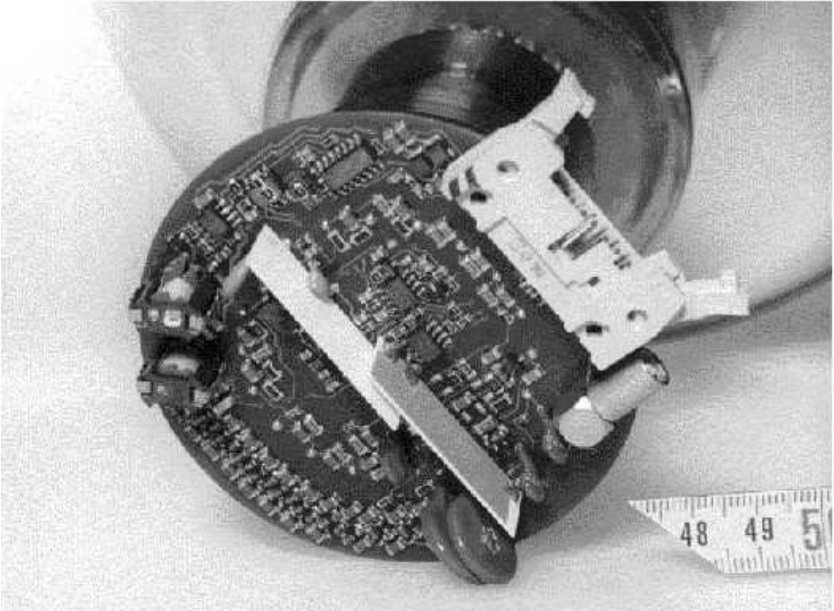,width=0.8\linewidth}}
 \end{center}
 \vspace{0cm}
 \caption{PMT base with integrated high voltage power supply.}
 \label{fig:iseg}
\end{figure}

A modified version of the base integrated HV supply PHQ5912 developed by iSeg\footnote{iSeg Technologies Germany GmbH, Munich Airport Center, Terminalstrass Mitte 18, 85356 Munich, Germany.} has been chosen. 
The HV and the distribution for the dynodes is 
generated on the base itself, which makes the design compact  
(figure~\ref{fig:iseg}) and requires only  low voltage supply ($\pm$5~V) 
and cabling. 
The Cockroft-Walton scheme is used to reduce power
consumption to less than 300~mW,  a factor 10 reduction compared to standard DC/DC converters and passive dividers. 

Voltage stability is guaranteed at the 10$^{-4}$ level. 
The voltages for the last 
3 dynodes are actively stabilized so as to remain constant even under
heavy load. The behaviour
at high pulse rates has been measured to be satisfactory for our purposes 
(gain variation below 5\% at 100~kHz). The recovery time after an intense 
burst of light (due, for example, to bioluminescence) is less than 0.5~s. 

\subsubsection{LED system}      

\begin{figure}
\vspace{-0cm}
 \begin{center}
 \mbox{\epsfig{file=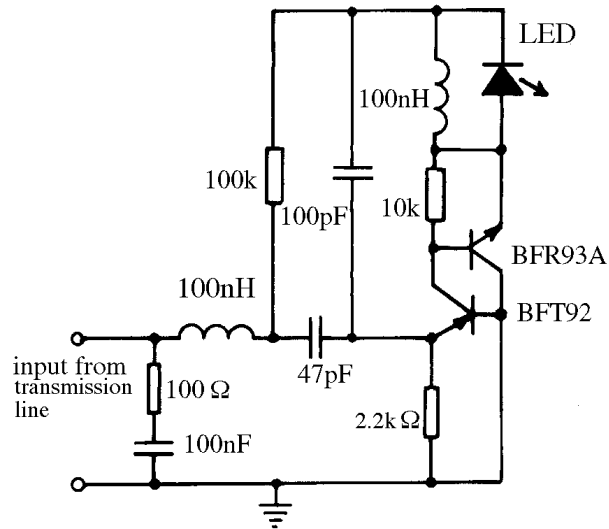,width=0.6\linewidth}}
 \end{center}
 \vspace{0cm}
 \caption{LED pulsing circuit.}
 \label{fig:pulser}
\end{figure}

Each OM is equipped with a LED system used for internal calibration 
(especially PMT transit time calibration). The system
is composed of a fast blue LED\footnote{HLMP-CB15 from 
Agilent technologies, 3500 Deer Creek Rd.,
Palo Alto, CA 94304, USA.} 
with peak intensity around 470~nm and
its pulser. The LED is glued on the back of the PMT in order to optimally 
illuminate
the centre of the photocathode through the aluminium coating of the tube.
The pulsing circuit, 
whose role is to deliver short current pulses (typically a few ns) 
is based on an original design from Kapustinsky
et al.~\cite{bib:pulser}, adapted for use with the most recent LEDs and 
optimized to reduce its electrical influence on nearby circuits (figure~\ref{fig:pulser}).   
Both the pulse amplitude and the trigger frequency can be adjusted from a single
control signal made of a triggering pulse superimposed on a DC level.
The main properties of the pulser are summarized in table~\ref{tab:pulser}.
The maximum light output has been measured to be of the order of 10$^8$
photons per pulse ($\sim$~40~pJ).

\begin{table}
\begin{center}
\caption{Main properties of the LED system.}
~\\
\begin{tabular}{ll} \hline
Pulse rise time & $\sim$ 2 ns \\ 
Overall pulse width & 4.5 -- 6.5 ns \\ 
Trigger-pulse jitter  & $\sim$ 100 ps \\ 
Trigger rate & 0 to 10 kHz \\ 
Light output (per pulse) & 0 to 40~pJ \\ 
Power consumption (@ 10kHz) & 0.3~mW \\ \hline
\end{tabular}
\label{tab:pulser}
\end{center}
\end{table}

\subsubsection{Electrical connections}
\label{sec:connectic}

The electrical connection between the OM and the local control module
(LCM) is made by a penetrator fixed in a hole in the glass envelope. 
Using a penetrator rather
than a connector offers more long term reliability.
The penetrator, provided by Euroceanique\footnote{Euroc\'eanique, 645 rue Mayor de Montricher, 13854 Aix en Provence, France.}, 
has  a diameter of 20~mm. The cable connecting the OM to the LCM 
is 1.9~m long and contains 
5 shielded twisted pairs of 0.4~mm$^2$ allocated as follows:
\begin{itemize}
\item One  for power transmission (Max. 50~V DC, 0.1~A)
\item Two  for the PMT signals (100~Ohms)
\item One for high voltage control and monitoring (0 to 4~V)
\item One for the LED trigger signal (0 to +24~V) 
\end{itemize}

In order to increase the dynamic range, the PMT anode and dynode 12 (D12)
signals  are both sent to the LCM where they are digitized. Differential transmission is used to reduce noise, with one twisted pair connected 
between the anode and D14, and the other between D12 and ground.

\subsection{OM assembly}
\label{ref:assembly}

\begin{figure}
\vspace{-0cm}
 \begin{center}
 \mbox{\epsfig{file=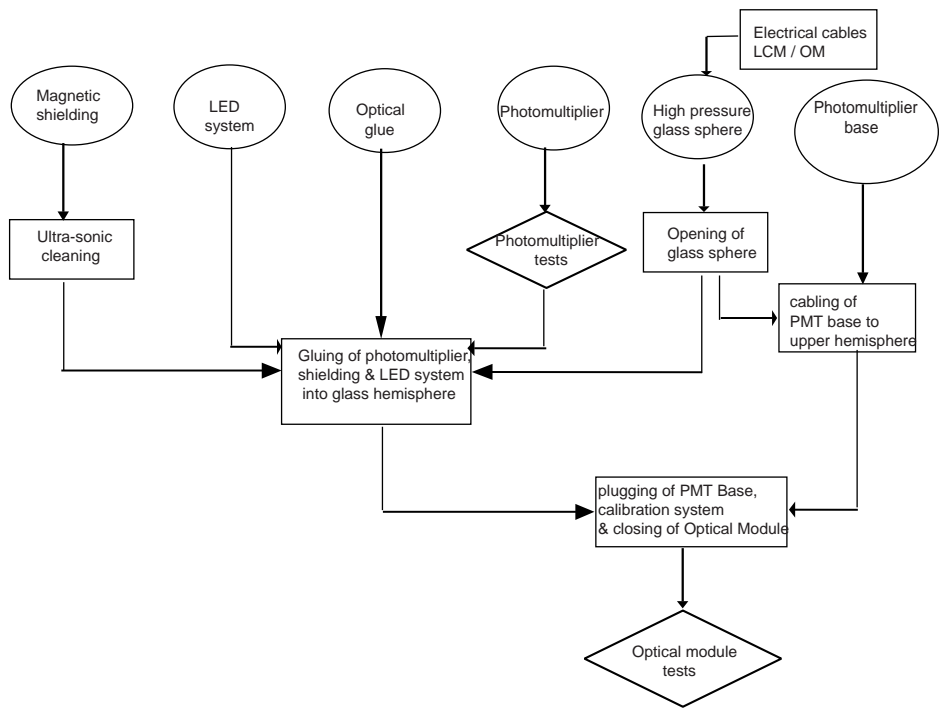,width=.9\linewidth}}
 \end{center}
 \vspace{0cm}
 \caption{OM assembly step by step.}
 \label{fig:assembly}
\end{figure}

The assembly of an OM from the components described above takes about 
8 hours (including tests) 
and consists of several steps (figure~\ref{fig:assembly}):

{\it 1) Opening of the high pressure sphere}\\
High pressure spheres are delivered closed with an absolute internal pressure of 800~mbars. Each sphere is equipped with a small manometer so as to easily monitor the internal pressure during the assembly process. 
The upper hemisphere is already painted black and equipped with its penetrator.
The sphere is opened by 
allowing air inside through the vacuum valve.

{\it 2) Cabling of the PMT base}  \\
The PMT base is connected by soldering to the twisted pairs of the penetrator 
contained in the upper  part of the glass sphere.

\begin{figure}
\vspace{-0cm}
 \begin{center}
  \mbox{\epsfig{file=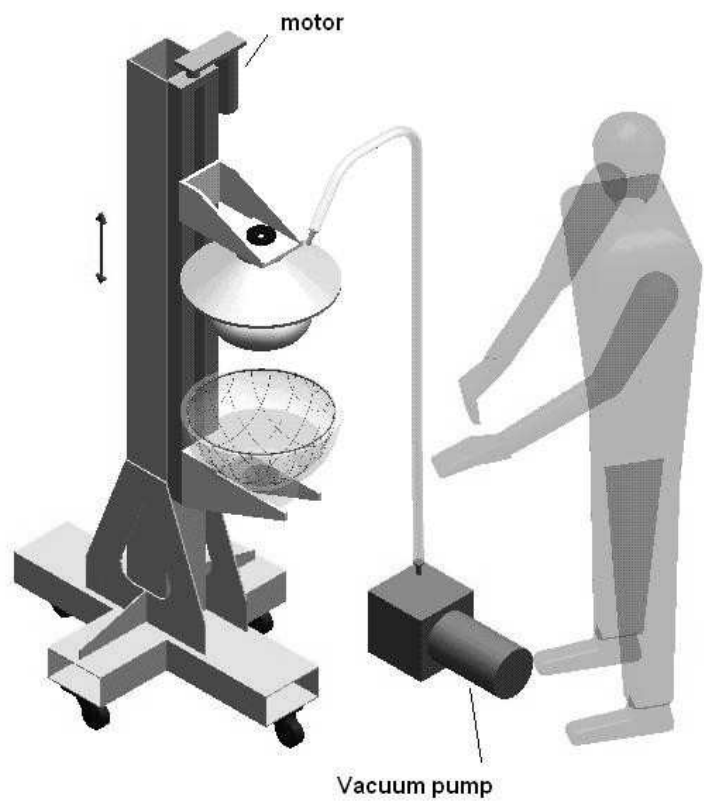,width=0.8\linewidth}}
 \end{center}
 \vspace{0cm}
 \caption{Schematic 3D view of the gluing bench.}
 \label{fig:gluing}
\end{figure}

{\it 3) Mounting of the PMT and $\mu$-metal cage}\\
The lower glass hemisphere is mounted on a gluing bench specifically de\-ve\-loped  (figure~\ref{fig:gluing}) which allows 
accurate (better than 1~mm)  
positioning of the PMT and the magnetic cage inside the glass sphere.

First, each element has to be carefully cleaned to avoid 
introducing impurities that could generate bubbles in the gel after polymerization.
The PMT glass window and the inner surface of the sphere are cleaned 
using methyl alcohol at room temperature. The $\mu$-metal cage is immersed for 15~min in an  ultrasonic bath of 
methyl alcohol at 40$^\circ$C.

The gluing is performed in three phases:
\begin{itemize}
\item First, 2 kg of
silicone gel is poured into the glass hemisphere and outgassed. 
 During this phase, many bubbles form and are evacuated, 
while the volume of the gel nearly doubles temporarily.
\item  
Next, the $\mu$-metal cage and the photomultiplier are positioned very slowly 
in the gel and the ensemble is outgassed at about 1~mbar absolute 
to remove the bubbles. The outgassing operation is
repeated three times for 3~minutes. The outgassing is limited to 
  9 minutes to avoid removing too much solvent. 
\item Finally, the gel is left to polymerize at room temperature and pressure for at least 4 hours.
\end{itemize}

{\it 4) Optical module closure}\\
The PMT base connected to the upper hemisphere is plugged onto the PMT and
the 2  hemispheres are aligned and  joined. 
The OM is then sealed by applying an under-pressure of 200~mbar 
inside the sphere and by using putty and tape
externally at the joint between the two hemispheres.


\section{OM characterization}
\label{sec:omtest}

Many tests were necessary to finalize the choice 
of individual components of the OM and in particular of 
the photomultiplier~\cite{bib:pm_ANTARES}. 
In addition, an evaluation of  the overall behaviour of the OM 
once all components are assembled  should  be performed,
both to verify the assembly procedure  and to estimate
the figure of merit of a particular design. Several test set-ups have been 
designed 
and implemented for this purpose and are presented in this section.

\subsection{Uniformity of the OM response} 

\begin{figure}
\vspace{-0cm}
 \begin{center}
 \mbox{\epsfig{file=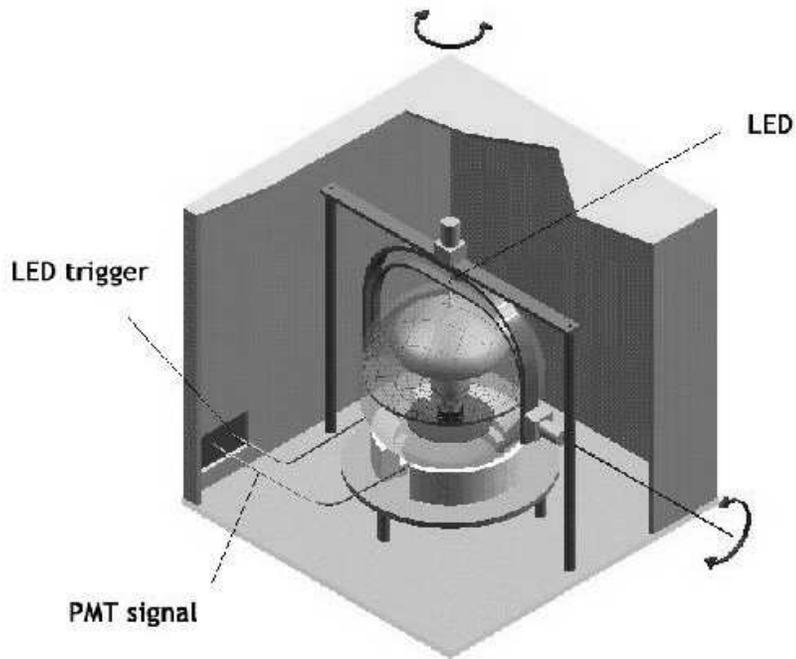,width=0.8\linewidth}}
 \end{center}
 \vspace{0cm}
 \caption{Schematic view of the scanning table used to measure the 
  uniformity of the OM response.}
 \label{fig:scanning_layout}
\end{figure}

 The aim of one of these measurements is to look for possible non-uniformities 
in the OM response
 by scanning the entire sensitive area with a point-like light source.
 The measurement is performed in the air, with the OM  located in a 
 dark box equipped with a mechanical 
 system allowing a  focussed  blue LED to scan 
the entire area  of the photocathode automatically 
(figure~\ref{fig:scanning_layout}). The LED always points to the center of the
OM and the light spot diameter is
 about 1~cm at the photocathode. 
 Non-uniformities may have several origins: asymmetrical PMT photocathode 
 or collection  efficiency, influence of the Earth magnetic field,
 defects (bubbles, dust) in the gel or glass sphere, etc. They might bias the
 detection efficiency of the telescope and they are very difficult to correct 
for {\it a posteriori}.

\begin{figure}
\vspace{-0cm}
 \begin{center}
 \mbox{\epsfig{file=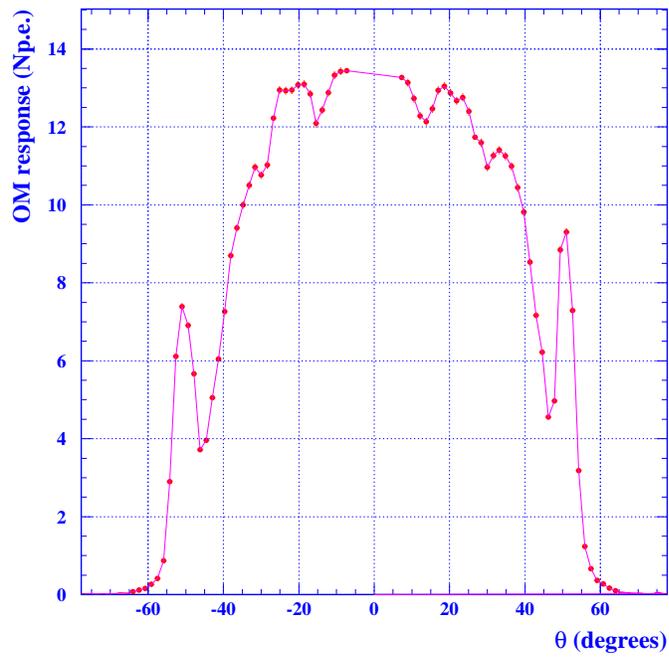,width=0.7\linewidth}}
 \end{center}
 \vspace{0cm}
 \caption{Result of the measurement of the OM uniformity response as
a function of the zenith angle $\vartheta$ for an arbitrary azimuthal angle.}
 \label{fig:scanning_res}
\end{figure}

The OM response (in number of photo-electrons) as a function 
of the position of the light spot is presented in 
figure~\ref{fig:scanning_res}. 
For central illumination ($\theta=0^{\circ}$) the OM
sees on average 13 photo-electrons and the OM response is essentially 
uniform up to $\pm 25^{\circ}$.
At $\theta=40^{\circ}$, the OM response drops quickly when the light 
spot reaches
the edge of the photocathode. The peaks at $\pm 50^{\circ}$ are 
 due to the gel-air dioptre which 
reflects an important fraction of the light back to the 
photocathode at this particular angle.

\subsection{Absolute calibration}

\begin{figure}
\vspace{-0cm}
 \begin{center}
 \mbox{\epsfig{file=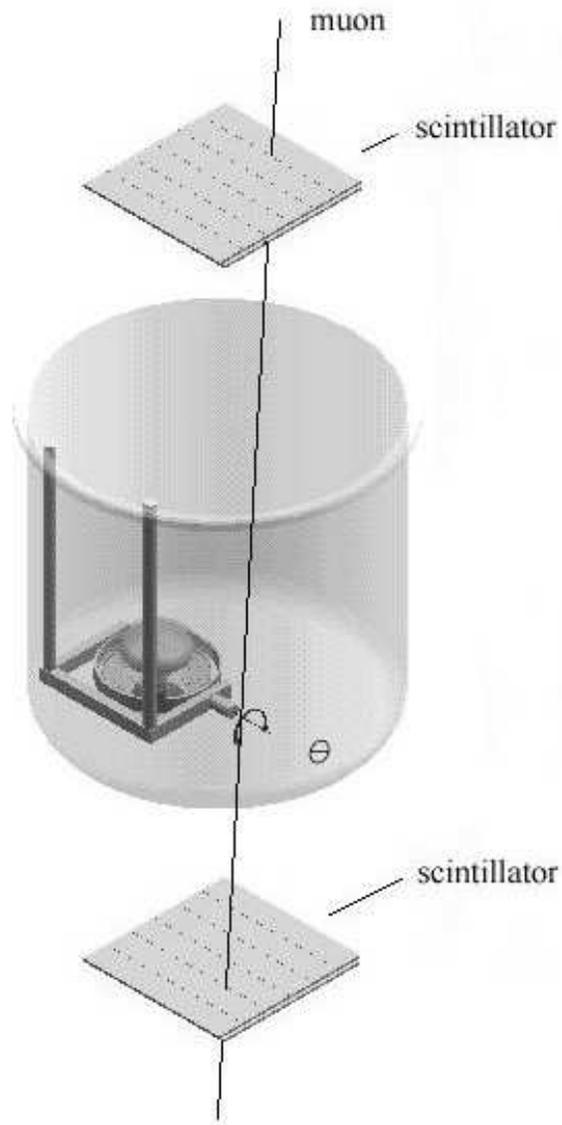,height=0.7\textheight}}
 \end{center}
 \vspace{0cm}
 \caption{Schematic view of the water tank used for absolute calibration
 of the OM. It is 1.5~m high, with a radius of 70~cm, and contains 2.3~m$^3$ of water kept
 pure by constant recycling. Above and below the tank, 3.6~m distant
from each other, are two crossed scintillators plane 
(6 scintillator hodoscopes of 12$\times$70~cm$^2$).}  
 \label{fig:gamelle_layout}
\end{figure}

 Absolute calibration and the determination of the overall response of the OM 
 are carried out in the laboratory using a water tank under light conditions
 similar to those found in the deep-sea environment: 
 a completely dark water volume, occasionally illuminated by flashes
 of \v{C}erenkov light produced by the passage of energetic muons.
 The set-up consists of a light-tight 
 cylindrical steel tank containing 2.3~m$^3$ of pure water 
(figure~\ref{fig:gamelle_layout}). Immersed in the tank about 1.2~m deep,
 the OM  detects \v{C}erenkov light pulses produced 
 in the water by nearly vertical cosmic rays. 
Above and below the steel tank, two pairs of scintillator hodoscopes
detect the passage of these particles. 
A trigger is formed by requiring a time coincidence on the 4 scintillator hodoscopes. Triggers are mainly due to muons with about 1 GeV energy.

The OM signal is sent to an ADC for digitization when 
an event trigger is received.
Two pattern units are also read out and stored 
for each event to determine which scintillators were hit. 
This information is used to reconstruct the position of the muon track
with respect to the centre of the OM. The two upper hodoscopes
determine the 
entry point of the muon, and the lower hodoscopes determine the exit point.
The horizontal coordinates ($x$ and $y$) of the entry and exit points are 
determined with a precision of about 6~cm. The
distance from the muon track to the centre of the OM can be
estimated with a precision of 2--3~cm.

A motor allows the OM to be rotated around a horizontal axis
passing through the centre of the sphere and perpendicular to the tank's axis.
This makes it possible to measure the optical module response at different
angles of incidence 
(angle between the direction pointed by the PMT in the OM (the OM axis)
 and the muon track).
 The zenith angle $\theta_{OM}$ of the OM axis can vary from $0^{\circ}$ (OM 
looking up) to $180^{\circ}$  (OM looking down).

\begin{figure}
\vspace{-0cm}
 \begin{center}
 \mbox{\epsfig{file=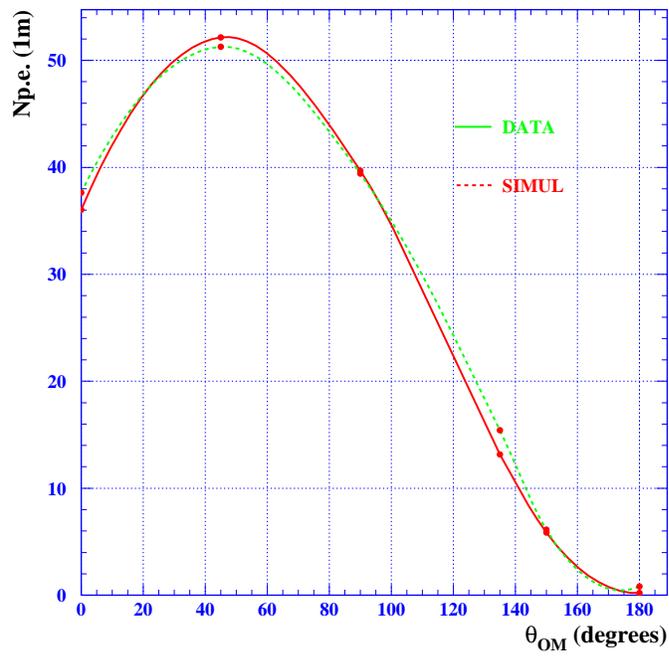,width=0.7\linewidth}}
 \end{center}
 \vspace{0cm}
 \caption{
Number of photo-electrons detected by the OM as a function of its zenith angle $\theta_{OM}$
for down-going atmospheric muons at 1~m distance. 
Data are compared to a detailed 
simulation based on GEANT3.21.}  
 \label{fig:gamelle_res}
\end{figure}

Figure~\ref{fig:gamelle_res} shows the response of an OM 
equipped with a 10$"$ R7081-20 Hamamatsu PMT operated at a gain of 
10$^8$ as a function of  $\theta_{OM}$. As shown in the figure,
for head-on illumination of the OM by the \v{C}erenkov cone 
($\theta_{OM}=42^{\circ}$), the OM gives an average of
52 photo-electrons for minimum ionizing muons at 1~m. The effective
angular acceptance of the OM, defined as the angular region where 
the OM sees more than half of the maximum signal, 
is $140^{\circ}$ for this head-on illumination.
Figure~\ref{fig:gamelle_res} also shows that the PMT 
is sensitive to a wide range of incident directions, extending up to
about 120$^{\circ}$  from its axis. This 120$^{\circ}$ limit 
was adopted for the design
of the  bottom part of the storey (figure~\ref{fig:omf}) 
to minimize the shadowing.

\subsection{Endurance test}

Experience has shown that an OM will undergo significant
environmental stresses over its lifetime, including:
\begin{itemize}
\item 
temperature variations during  storage and  transportation: --10$^\circ$C to +45$^\circ$C;
\item 
 mechanical stresses due to ground and sea transportation;
\item 
light exposure (a maximum of 1000~W/m$^2$ continuously for 8 hours
 from the sun light on the boat deck);
\item 
permanent contact with sea water during normal operation;
\item 
an external working pressure of 256 bars, and 307 bars for 1 hour during 
the acceptance tests.
\end{itemize}

To check whether the OM can  be subjected to all of these conditions without 
significant degradation, a 
test campaign for environmental qualification has been conducted.
To ensure reproducibility, 4 identical OMs have been built 
and have been subjected to 3 sets of tests: climatic, mechanical, and high pressure tests (table~\ref{tab:testenviron}).
At each step of the qualification procedure, the integrity of the OM is 
monitored by:
\begin{itemize}
\item visual inspection
\item measurement of the internal pressure 
\item high pressure tests 
\item measurement of the PMT response
\end{itemize}

All four OMs successfully passed these tests.

\begin{table}
 \begin{center}
\caption{Summary of endurance tests.}
\begin{tabular}{ll} \hline
{\bf Climatic tests }&   \\ \hline
Cold storage    & --10$^0$C for 72~h \\
Humid heat     & +50$^0$C and 93\% relative humidity for 96~h \\
Dry heat        & +50$^0$C for 16~h \\
Thermal shocks  & +50$^0$C in air / +15$^0$ in water in 10~s \\
Sun Exposure    & +50$^0$C and light flux of 1000 W/m$^2$ \\ \hline
{\bf Mechanical tests }& \\ \hline
Vibrations      & 1~mm from 5 to 16 Hz / 1~$g$ from 16 to 55~Hz \\
Shocks          & 100 half sine shocks of 15~$g$/20~ms \\ \hline
{\bf High pressure tests }&  307 bars for 1 h \\ \hline  
\end{tabular}
\label{tab:testenviron}
\end{center}
\end{table}

\subsection{In situ tests and long term behaviour}

Since the beginning of the ANTARES project, more than 30 OMs  have been 
built and used for deep-sea measurements for the site 
evaluation campaign. These OMs have been immersed for
periods ranging from a few hours  to more than
one year~\cite{bib:nathalie} without a single failure.
This  indicates that the OM fabrication is 
under control and is adequate for the intended purpose.

In November 1999, a prototype line connected to the shore via an electro-optical
cable was deployed 40~km of the coast of Marseilles at a depth of 1200~m. 
This line was equipped with 8 OMs very similar to the final OM design.
The OMs were 
stored for nearly 2 years before immersion, part of the
time outdoors  with minimum protection. The line 
was operated for about 6 months and recovered in June 2000.
All 8 OMs were tested after recovery and  were found to work as expected.

Tests were also performed in situ to quantify the destructive effect
of the implosion of one OM  in a storey at various depths. 
Storeys equipped with three empty spheres, one of which had been artificially 
weakened, were used. 
The implosion pressure could be varied
by grinding out a variable flat section on the outer surface, 
The tests showed that the damage depended on the depth, as expected. 
At 1900~m, the weakened sphere imploded without damaging the other two spheres or the
mechanical structure of the storey. But at 2600~m, the implosion of the weakened sphere triggered 
the implosion of the two other spheres and significantly distorted 
the mechanical structure. At this depth, the amount of energy released by the implosion 
is estimated to be in excess of 10$^6$ Joules.
Nonetheless, even when the three spheres of a 
storey are destroyed (the worst case), 
the integrity of the string itself is not affected.
It has to be emphasized that the failure of a sphere is most likely 
to result from  inappropriate handling during deployment and transportation, a 
risk that can be minimized by using special procedures at these stages.

\section{Mass production and quality control}
\label{sec:massprod}

For mass production, each element of the OM is supplied fully tested
and characterized by the different suppliers. 
Nevertheless, several measurements are foreseen at 
differents stages of the assembly chain to 
ensure the quality and reliability of the production.
 
Some systematic checks applied to all OMs allow 
the condition and basic functioning of each element to be assessed
in a quick and easy way:
\begin{itemize}
\item the inside pressure of the glass sphere is read 
 from its internal manometer upon reception and should be less than 
  900 mbars before opening;
\item the photomultiplier, which has already undergone a detailed acceptance 
  check, is tested with its final base just before gluing 
 simply by switching the HV on and recording the signal shape at nominal gain
 on a digital scope;
\item the elasticity of the optical glue is measured 24 hours after polymerization and should be in a predefined range (samples of each optical glue preparation are stored for future measurements);
\item the photomultiplier signal shape is recorded again just after closure 
 of the OM and should be  identical to the signal shape before gluing;
\item the inside pressure of the OM is measured again before storage and 
should be the same as that obtained after closing.
\end{itemize}   

If the OM fails to pass one of these tests during assembly, 
it is simply left out of the production chain.

More detailed checks are made by sampling only. 
They will be performed on about 5\% of all OMs,  chosen randomly.
 They consist of:
\begin{itemize}
\item a measurement of the refractive index of the optical glue on
 samples of the preparation  taken after polymerization; 
\item a high pressure test at 307 bars; 
\item a measurement of the OM response to short light pulses in a dedicated black box.
\end{itemize}

\section{Conclusion}

The optical module is a key component of the ANTARES detector.  
In view of its importance, extensive R\&D studies have been carried out 
by the collaboration to ensure that 
it would not only be optimized in terms of performance, but would 
also be robust enough
to withstand sea-operation stresses and be highly reliable so as
to guarantee proper operation during the entire lifetime of the detector.  
The general design and the choice of components have been 
finalized, so that mass production of the OMs can begin.


\section{Acknowledgements}

The authors acknowledge financial support by the funding agencies, 
in particular:
\
Commissariat \`a l'Energie Atomique, Centre Nationale de la Recherche Scientifique, 
Commission Europ\'eenne (FEDER fund), D\'epartement du Var and
R\'egion Provence Alpes C\^ote d'Azur, City of La Seyne, France; 
the Ministerio de Ciencia y Tecnolog\'{\i}a, Spain (FPA2000-1788);
the Instituto Nazionale di Fisica Nucleare, Italy;
the Russian Foundation for Basic Research, grant no. 00-15-96584, Russia; 
the foundation for fundamental research on matter FOM and the national 
scientific research organization NWO, The Netherlands.


\clearpage

\end{document}